\documentclass[12pt]{article}
\pdfoutput=1

\usepackage{cite}
\usepackage{epsfig}
\usepackage{amsmath}
\usepackage{amssymb}
\usepackage{cancel}
\usepackage{pstricks}
\usepackage{afterpage}
\usepackage{color}
\usepackage{slashed}
\usepackage{caption}
\usepackage{braket}
\usepackage{mathtools}
\usepackage{bbold}

\newcommand{\gs}{g_{\rm s}}
\newcommand{\as}{\alpha_{\rm s}}
\def\mathswitch#1{\relax\ifmmode#1\else$#1$\fi}
\newcommand{\msbar}{\mathswitch{\overline{\text{MS}}}\ }
\newcommand{\tev}{\,\, \mathrm{TeV}}

\newcommand{\mycaption}[1]{\caption{\sl #1}}
\renewcommand{\thefootnote}{\roman{footnote}} 
\renewcommand{\thefootnote}{\fnsymbol{footnote}}

\makeatletter
\def\section{\@startsection {section}{1}{\z@}{+3.0ex plus +1ex minus
  +.2ex}{2.3ex plus .2ex}{\large\bf\boldmath}}
\def\subsection{\@startsection{subsection}{2}{\z@}{+2.5ex plus +1ex
minus +.2ex}{1.5ex plus .2ex}{\normalsize\bf\boldmath}}
\def\subsubsection{\@startsection{subsubsection}{3}{\z@}{+3.25ex plus
 +1ex minus +.2ex}{1.5ex plus .2ex}{\normalsize\it}}

\graphicspath{{.}{FIGURES/}{FIGURES2/}}

\begin{document}
\thispagestyle{empty}

\def\thefootnote{\fnsymbol{footnote}}

\begin{flushright}
\end{flushright}

\vspace{1cm}

\begin{center}

{\Large {\bf Renormalization and ultraviolet sensitivity of gauge vertices
in universal extra dimensions}}
\\[3.5em]
{\large
Ayres~Freitas and Daniel Wiegand
}

\vspace*{1cm}

{\sl\noindent
Pittsburgh Particle-physics Astro-physics \& Cosmology Center
(PITT-PACC),\\ Department of Physics \& Astronomy, University of Pittsburgh,
Pittsburgh, PA 15260, USA
}

\end{center}

\vspace*{2.5cm}

\begin{abstract}
When computing radiative corrections in models with compactified extra
dimensions, one has to sum over the entire tower of Kaluza-Klein excitations
inside the loops. The loop corrections generate a difference between the
coupling strength of a zero-mode gauge boson and the coupling strength of its
Kaluza-Klein excitation, although both originate from the same
higher-dimensional gauge interaction. Furthermore, this discrepancy will in
general depend on the cutoff scale and assumptions about the UV completion
of the extra-dimensional theory. In this article, these effects are studied in
detail within the context of the minimal universal extra dimension model (MUED). The
broad features of the cutoff scale dependence can be captured through the
solution of the functional flow equation in five-dimensional space. However, an
explicit diagrammatic calculation reveals some modifications due to the
compactification of the extra dimension. Nevertheless, when imposing a physical
renormalization condition, one finds that the UV sensitivity of the effective
Kaluza-Klein gauge-boson vertex is relatively small and not very important for
most phenomenological purposes. Similar conclusions should hold in a
larger class of extra-dimensional models besides MUED.
\end{abstract}

\setcounter{page}{0}
\setcounter{footnote}{0}

\newpage


\section{Introduction}

\noindent 
In models with additional spatial dimensions, the Standard Model (SM) gauge
dynamics are the low-energy remnant of a gauge theory in the full
higher-dimensional space. Let us consider one flat extra dimension compactified
on a circle with radius $R$ and a non-Abelian gauge group ${\cal G}$ with gauge
coupling $g^{(5)}$ in the five-dimensional (5D) theory. At tree level, there is
a simple geometric relation between $g^{(5)}$ and the coupling $g$ in the 4D
compactified description, which is given by $g^{(5)} = g/\sqrt{\pi R}$.

At loop level, however, the 4D and 5D couplings develop a non-trivial dependence
on the renormalization and cutoff scales
\cite{ued,Dienes:1998vh,Dienes:1998vg}. This is intimately tied to the fact that
extra-dimensional theories are superficially non-renormalizable. 
Moreover, in the compactified 4D
picture, the couplings of the zero-mode gauge bosons and the Kaluza-Klein (KK) gauge bosons
require different counterterms at one-loop order
\cite{Chivukula:2011ng,kkgluon}, although they both stem from the same 5D gauge
coupling.
This implies that these two 4D couplings, while identical at tree-level, differ
by loop-induced correction terms. In fact, one also has to entertain the
possibility that  this correction may depend on assumptions about the nature of
the UV completion of the 5D theory.

While the question of scale sensitivity of couplings in extra-dimensional
theories has a long history
\cite{ued,Dienes:1998vh,Dienes:1998vg,original,other,Gies:2003ic,Rothstein:2003mp},
a detailed analysis of the \emph{difference} between the effective zero-mode
gauge bosons and their KK excitations is so far missing. Such an analysis is the
topic of this paper, together with an extensive discussion of the influence of
the UV cutoff scale on both vertices.

As a concrete example, the decay $G_1 \to Q_1 \bar{q}$ in the minimal universal
extra dimension model (MUED) is considered, where $G_1$ and $Q_1$ are a level-1
KK gluon and quark, respectively, while $q$ is a SM quark of the same flavor.
This decay is kinematically allowed due to the loop-induced mass corrections to
the KK-gluon and -quark \cite{cms,fkw}. The complete set of one-loop QCD
corrections in MUED can be grouped into two parts: (a) diagrams involving
only zero and level-1 modes in the loops, and (b) diagrams involving KK
modes beyond level~1. Category (a) contains all IR divergent contributions,
which cancel against corresponding real emission corrections. It is equivalent
to the corrections within a low-energy effective theory dubbed the two-site
coloron model \cite{kkgluon}. Category (b) contains the dependence on the
cutoff scale and the renormalization procedure
. 
In order to provide a physical renormalization condition for the $G_1Q_1\bar{q}$
vertex, it must be compared to the  $gq\bar{q}$ vertex in MUED, which will be
matched to the SM  below the compactification scale. Here $g$ is the SM gluon
field.

While the 5D gauge coupling does not run within the usual framework of
dimensional regularization, its dependence on a cutoff scale can be described by
the functional flow equation \cite{Wettericheq,bkgfield}. It is very
illustrative to compare the flow equation analysis to the explicit summation
over the KK tower, as will be shown below.

After a brief review of the conventions and notation for the MUED and
coloron model in section~\ref{sc:mod}, the calculation of the next-to-leading
order (NLO) corrections to the decay $G_1 \to Q_1
\bar{q}$ in the coloron model is described in section~\ref{sc:dec}. This set of
corrections corresponds to the category (a) above. The contribution from
level-$n$ KK modes ($n>1$) in the loops is discussed in section~\ref{sc:ued}.
Special emphasis is given to the dependence on the cutoff scale in different
approaches (explicit Feynman diagram calculation, large-$n$ approximation,
functional flow equation). Finally, section~\ref{sc:pheno} illustrates the
impact of the cutoff dependence on physical quantities, such as the decay $G_1 \to Q_1
\bar{q}$, before the conclusions are presented in section~\ref{sc:concl}.


\section{QCD sector of MUED and Coloron Model}
\label{sc:mod}

\noindent
In this section, a brief overview of the QCD sector of MUED and the two-site coloron
model are given, to clarify the notation and conventions.

The Lagrangian for the strongly interacting MUED fields can be written as
\begin{equation}
{\cal L}_{\rm MUED,QCD} = {\cal L}_{\rm gauge} + {\cal L}_{\rm gf} + {\cal L}_{\rm
ghost} + {\cal L}_{\rm quark}.
\end{equation}
Here ${\cal L}_{\rm gauge}$ contains the kinetic term of the 5D gluon field. For
practical purposes, one also needs to introduce a gauge-fixing term, ${\cal
L}_{\rm gf}$, and corresponding ghost Lagrangian, ${\cal L}_{\rm ghost}$. In
Feynman gauge these contributions read
\begin{align}
{\cal L}_{\rm gauge} &= \frac{1}{2} \int_{-\pi R}^{\pi R} dx^5 \, 
 \Bigl [ -\frac{1}{4} G^a_{MN} G^{a,MN} \Bigr ], \\
{\cal L}_{\rm gf} &= \frac{1}{2} \int_{-\pi R}^{\pi R} dx^5 \,
 \Bigl [ -\frac{1}{2} (\partial^M G_M^a)^2 \Bigr ], \\
{\cal L}_{\rm ghost} &= \frac{1}{2} \int_{-\pi R}^{\pi R} dx^5 \,
 \Bigl [ -\bar{c}^a \partial^M \partial_M c^a - \gs^{(5)} f^{abc}
 (\partial^M \bar{c}^a) G_M^c c^b \Bigr ].
\end{align}
Here $x^5$ is the coordinate of the extra dimension, which is compactified on a
circle with radius $R$. 
Furthermore, $G^a_{MN}$ is the field strength tensor of the 5D gluon field
$V^a_M$, while $c^a$ denotes the ghost field. 
Capital Latin indices run over all five dimensions ($M,N = 0,1,2,3,5$), whereas
Greek indices ($\mu,\nu,...$) run over the usual uncompactified four dimensions.
The superscripts $a,b,...$ refer to SU(3) indices in the adjoint representation,
whereas $\gs^{(5)}$ refers to the 5D QCD coupling, which is related to the 4D
coupling according to $\gs^{(5)} = \gs/\sqrt{\pi R}$.

The coupling of the gluon field to quarks is described by
\begin{equation}
{\cal L}_{\rm quark} = \frac{1}{2} \int_{-\pi R}^{\pi R} dx^5 \, 
 \bigl [ \overline{\Psi} \, \Gamma^M (\partial_M + i\gs^{(5)}G_M^a T^a) \Psi \bigr
 ],
\end{equation}
where $T^a$ are the SU(3) generators in the fundamental representation, and
$\Psi$/$\Psi'$ refer to the quark fields that are doublets/singlets under SU(2),
respectively. Upon
compactification and orbifolding, the 5D fields decompose into towers of 4D KK
excitations:
\begin{align}
G^{a,\mu}(x,x^5) &= \frac{1}{\sqrt{\pi R}} \biggl [ g^{a,\mu}(x) + \sqrt{2}
 \sum_{n=1}^\infty G_n^{a,\mu}(x) \, \cos \frac{n x^5}{R} \biggr ], \\
G^{a,5}(x,x^5) &= \sqrt{\frac{2}{\pi R}} \biggl [ 
 \sum_{n=1}^\infty G_n^{a,5}(x) \, \sin \frac{n x^5}{R} \biggr ], \\
c^a(x,x^5) &= \frac{1}{\sqrt{\pi R}} \biggl [ c^a(x) + \sqrt{2}
 \sum_{n=1}^\infty c_n^a(x) \, \cos \frac{n x^5}{R} \biggr ], \\
\Psi(x,x^5) &= \frac{1}{\sqrt{\pi R}} \biggl [ P_- q_L(x) + \sqrt{2}
 \sum_{n=1}^\infty \Bigl ( P_- Q_{n,L}(x) \, \cos \frac{n x^5}{R} +
 P_+ Q_{n,R}(x) \, \sin \frac{n x^5}{R} \Bigr ) \biggr ], \\
\Psi'(x,x^5) &= \frac{1}{\sqrt{\pi R}} \biggl [ P_+ q_R(x) + \sqrt{2}
 \sum_{n=1}^\infty \Bigl ( P_- Q'_{n,L}(x) \, \cos \frac{n x^5}{R} +
 P_+ Q'_{n,R}(x) \, \sin \frac{n x^5}{R} \Bigr ) \biggr ],
\end{align}
with $P_\pm = (1\pm \gamma^5)/2$. Notice that the Dirac fermions $Q_n = Q_{n,L} + Q_{n,R}$ are constructed from components with the same SU(2)$\times$U(1) quantum numbers, while the left and right components of $q = q_L+q_R$ come from the 5D doublet and singlet respectively.\footnote{We will be dealing exclusively with QCD processes in this paper, which produce identical results for doublet and singlet quarks $Q$ and $Q'$ as external legs. For concreteness we will use $Q$ in the discussion.}
The explicit Lagrangian terms in the 4D compactified form can be found for
example in Ref.~\cite{fkw}.

\medskip

The two-site coloron model described in Ref.~\cite{kkgluon} has the same degrees
of freedom as zero and level-1 modes of the strongly interacting sector of MUED.
It is based on a SU(3)$\times$SU(3) symmetry broken by a non-linear sigma model,
which ensures that gauge symmetry is conserved. This is in contrast to a naive
truncation of MUED at the KK level 1, which violates the gauge symmetry of the
5D gluon field \cite{uedgv}. A detailed description of the model can be found in
Ref.~\cite{kkgluon}. 

For the purposes of this work it suffices to point out that the only difference
in the Feynman rules of the coloron model and MUED occurs for quadruple
couplings of four level-1 bosons. Such couplings do not contribute to the
calculation of level-1 KK gluon decays at LO and NLO. Therefore, the result for these
decays in the coloron model is identical to what one would obtain in the naively
truncated MUED\footnote{One should keep in mind, however, that this is a process-dependent
statement, and naively truncated MUED is not a well-defined approximation for
other processes involving KK gluons, such as KK-gluon pair production
\cite{kkgluon}.}.


\section{KK-gluon decay in Coloron Model}
\label{sc:dec}

To account for the IR physics of the corrections to the $G_1Q_1\bar{q}$ vertex we calculate the one-loop decay width $\Gamma[G_1 \rightarrow \bar{q}\, Q_1]$ within the coloron model. As already mentioned above, a na\"ively truncated MUED model violates gauge invariance \cite{uedgv}, whereas the coloron model can be used as a gauge-invariant low-energy approximation of MUED.

\begin{figure}[t]
\vspace{1em}
\centering
\captionsetup{width=\linewidth}
\includegraphics[width=5in]{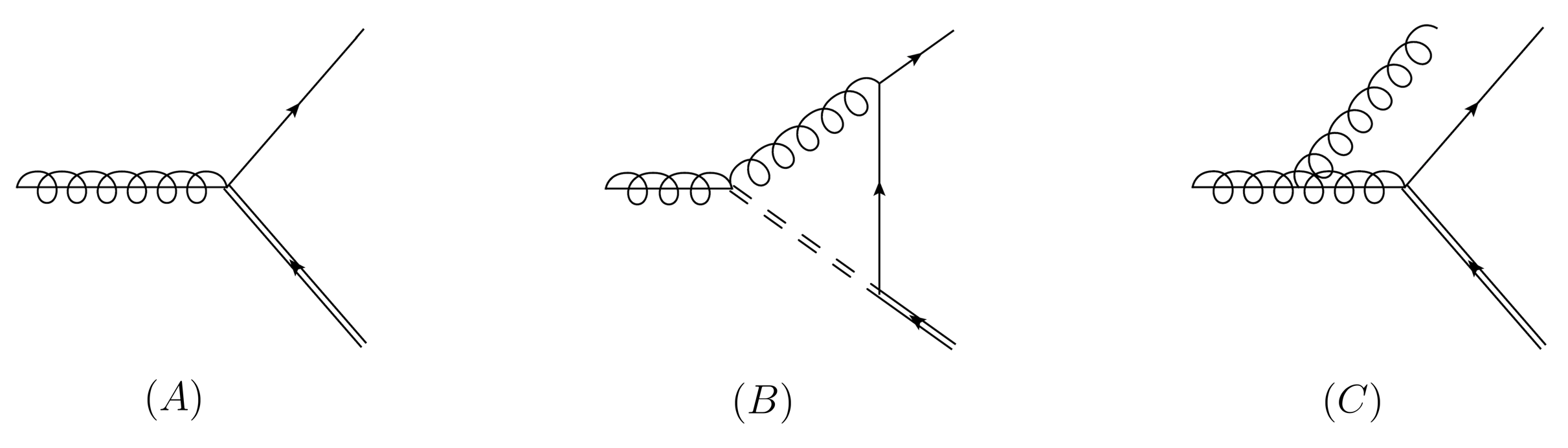}
\vspace{-1ex}
\mycaption{Sample diagrams the decay of the level-1 gluon decay $G_1 \rightarrow q \overline{Q}_1$ within the coloron model. (A) Born-level decay; (B) NLO correction to the born vertex; and (C) real emission contribution with an infrared divergence. Here a spring--straight line, solid double line and broken double line indicate the propagator of a coloron, heavy quark, and Goldstone mode of the coloron, respectively.
\label{fig:ColoronDecaydiag}}
\end{figure}

Diagram (A) in Fig.~\ref{fig:ColoronDecaydiag} represents the only tree-level Feynman describing the coloron decay. The squared matrix element at leading order is given by
\begin{align}
|{\cal M_\textrm{Born}}|^2 = \frac{g^2}{16M_{G_1}^2}N_CC_F\left(M_{G_1}^2-M_{Q_1}^2\right)\left(2M_{G_1}^2+M_{Q_1}^2\right).
\label{eq:MBorn}
\end{align}
where $M_{G_1}$ and $M_{Q_1}$ are the masses of the coloron and the heavy quark, respectively. When matching the coloron model to MUED at leading order, one obtains the constraint $M_{G_1} = M_{Q_1} = R^{-1}$, so that \eqref{eq:MBorn} vanishes. We therefore assume the radiatively induced mass splitting as in MUED, amounting to \cite{fkw}
\begin{align}
\frac{M_{G_1}^2-M_{Q_1}^2}{R^{-1}} = \sqrt{1+\frac{\gs^2}{96\pi^2}\left(154+69\log{\Lambda R}-\frac{9\zeta(3)}{\pi^2}\right)}-1-\frac{\gs^2}{48\pi^2}(16+9\log{\Lambda R})
\,,
\end{align}
where $\Lambda$ is the high-energy cutoff scale of MUED.\\
Since the mass splitting is small in comparison to the mass scale of $G_1$ and $Q_1$, it is sufficient to perform the calculation of the NLO corrected matrix element only to lowest order in $M^2_{G_1}-M^2_{Q_1}$, i.e. the mass splitting can be ignored inside the one-loop integrals.

\bigskip

The phase space integration for the two-body process can be performed analytically and leads to the decay width
\begin{align}
\Gamma_2\left[G_1 \rightarrow \bar{q}\, Q_1\right] = \frac{M_{G_1}^2 - M_{Q_1}^2 }{16\pi M^3_{G_1}}|{\cal M}|^2.
\end{align}
The renormalization is performed by using the on-shell scheme for the wave-function renormalization of the physical states and \msbar renormalization for the strong coupling constant. An example loop diagram is shown in Fig.~\ref{fig:ColoronDecaydiag}~(B).

For the external states the wave function renormalization constants
\begin{align}
\delta Z^\psi_{\rm L} = 
\delta Z^\psi_{\rm R} \quad [\psi=q,q'], \qquad
\delta Z^{Q_1}_{L,R} = \delta Z^{Q'_1}_{R,L}, \qquad
\delta Z^C,
\end{align}
are introduced for the left- and right-handed (massless) SM quarks, the doublet and singlet KK-quarks and the coloron, respectively. As usual, their values are determined through the residues of the renormalized propagators, leading to 
\begin{align}
\delta Z^\psi_{\rm L,R} &= -{\Re\rm e}\{ \Sigma^\psi_{\rm L,R}(0)\}, &
\delta Z^C &= -{\Re\rm e}\bigl\{ \tfrac{\partial}{\partial(p^2)}\Sigma^C(M_{G_1}^2)
\bigr\},
\end{align}
and
\begin{align}
\delta Z^Q_\textrm{L,R} = -{\Re\rm e}\left\{\Sigma^Q_\textrm{L,R}(M^2_{Q_1})\right\} - M^2_{Q_1}\frac{\partial}{\partial p^2}{\Re\rm e}\left\{\Sigma^Q_\textrm{R}(p^2)+\Sigma^Q_\textrm{L}(p^2)+2\Sigma^Q_\textrm{S}(p^2)\right\}\bigg|_{p^2=M^2_{Q_1}}.
\end{align}
We use $\Sigma_{\rm L,R}(p^2)$ and $\Sigma^C(p^2)$ to denote the left/right-handed quark self-energies and the transverse coloron self-energy, respectively.
Furthermore we renormalize the coupling constant in the \msbar scheme, leading to a redefinition of the coupling according to  
\begin{align}
&\gs^{\rm bare} \to \gs(\mu)\,\bigl (1+ \delta Z_g \bigr ) \\
&\delta Z_g =  \frac{\as(\mu)}{4\pi} \biggl [-\frac{\beta_\textrm{C}}{2} \biggl ( \frac{1}{\epsilon} - \gamma_{\rm E} + \log(4\pi)
  \biggr) + \frac{\beta_0}{2}\log{\frac{R^{-2}}{\mu^2}}\biggr ],
  \label{eq:Vertexrenorm}
\end{align}
where the beta functions can be written as
\begin{align}
\beta_\textrm{C} = \Bigl (\frac{(3+85C_A^2)(C_A-2C_F)}{12}-
 \frac{8}{3}n_qT_f \Bigr ), \quad \beta_0 = \Bigl (\frac{11}{3}C_A - \frac{4}{3}n_qT_f \Bigr ).
\end{align}
Note that for our analysis we take all SM quarks to be massless, including the top, such that $n_q = 6$.
The scale-dependence of the coupling is determined by fixing its value to the SM value of the strong coupling at $\mu = R^{-1}$, even though their counterterms are differing. Beneath the threshold $\mu = R^{-1}$ we assume the coupling to effectively not run at all. This is achieved through the logarithm term in \eqref{eq:Vertexrenorm}, which cancels the leading-order running of $\as(\mu)$ for $\mu < R^{-1}$. A similar decoupling prescription was used previously in Ref.~\cite{Beenakker:1997ut} in the context of supersymmetry.

\bigskip

To remove the infrared divergencies we employ phase-space slicing with two cutoffs as it is described in Ref.~\cite{Harris:2001sx}. The soft-collinear divergencies from the real emission diagrams, see Fig.~\ref{fig:ColoronDecaydiag}~(C) for an example, exactly cancel the soft poles of the one-loop functions. To analytically extract and cancel the poles the matrix element and phase space of the three-body decay are expanded for small emission angles $\theta$ and small gluon energies $E_3$. This approximation is used in the region bounded by the soft-collinear conditions
\begin{align}
&\cos{\theta} \geq 1-\frac{R^{-1}}{E_3}\delta_c\,, &
&\frac{R^{-1}}{2}\delta_s \leq E_3\,,
\end{align}
with soft and collinear cutoff parameters $\delta_s$ and $\delta_c$ respectively.

Both the matrix element and the phase space of the three-body decay factorize in the limit of small gluon energies into the two-body result and an eikonal factor
\begin{align}
\Gamma_\textrm{NLO}^S = \Gamma_\textrm{LO} \times
\frac{\as}{2\pi} \, \frac{\Gamma(1-\epsilon)}{\Gamma(1-2\epsilon)}\Bigl (\frac{4\pi\mu_{\rm R}^2}{R^{-1}}\Bigr )^\epsilon
\sum_{i,j}\int dS_g\, \frac{-p_i\cdot p_j}{(p_i \cdot p_g)(p_j \cdot p_g)}\,.
\end{align}
The remaining angular integrals can be performed analytically in dimensional regularization and are found in the literature, e.g.\ in Ref.~\cite{Beenakker:1988bq}.

Hard-collinear divergencies only arise from the diagram with a final-state gluon being emitted from the SM quark leg. The phase space and the three-body matrix element factorize in the limit of a small angle $\theta$ between quark and gluon into the Born contribution and a divergent Altarelli-Parisi splitting kernel. In dimensional regularization one finds
\begin{align}
\Gamma_\textrm{NLO}^C = \Gamma_\textrm{LO} \times
\frac{\as}{2\pi} \, \frac{\Gamma(1-\epsilon)}{\Gamma(1-2\epsilon)}\Bigl (\frac{4\pi\mu_{\rm R}^2}{R^{-1}}\Bigr )^\epsilon\left(\frac{A_1^c}{\epsilon}+A_0^c\right),
\end{align}
where $A_1^c$ and $A_0^c$ are numerical constants found e.g.\ in \cite{Harris:2001sx}.

\begin{figure}[tbp]
\centering
\captionsetup{width=\linewidth}
\vspace{-1em}
\includegraphics[width=12cm]{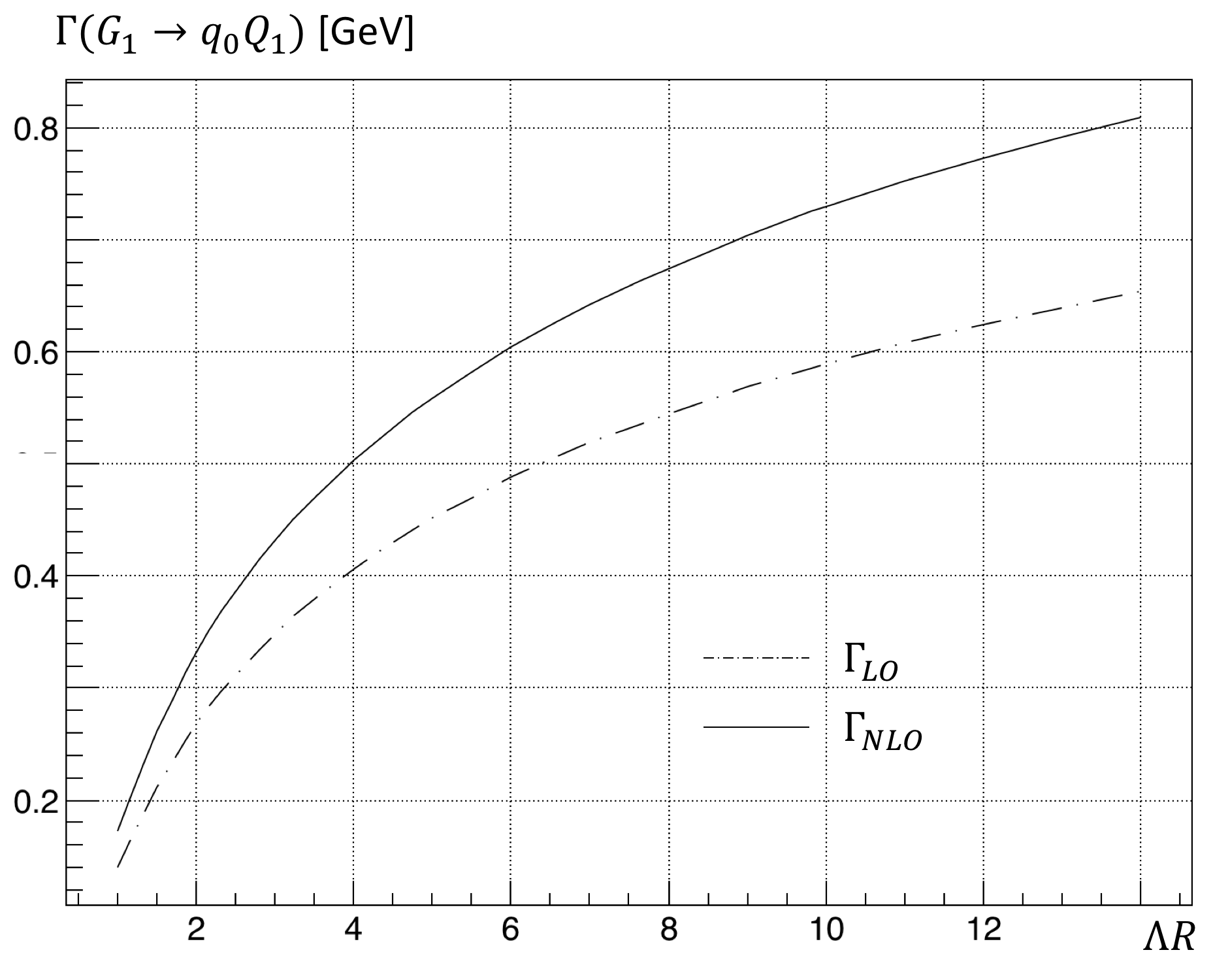}
\vspace{-1ex}
\mycaption{Full LO and NLO decay width for the level-1 KK-gluon decaying into a regular quark and the level-1 KK-quark within the Coloron model. Both curves exhibit the same $\Lambda$ dependence, since the $\kappa$-factor is constant in our approximation. We assume $R^{-1} = 1 \tev$}.
\label{fig:ColoronDecayWidth}
\end{figure}

The three-body phase space integration in the hard, non-collinear regime can be performed analytically, within the bounds set by the soft-collinear conditions. After carrying out the integration over the three-body matrix element the cancellation of cutoff parameters between the three and two-body processes is then checked explicitly. Due to the approximation $M^2_{G_1}-M^2_{Q_1} \ll M^2_{G_1}$ the ratio $\kappa \equiv \Gamma_\textrm{NLO}/\Gamma_\textrm{LO}$ is a constant, independent of any of the masses, and turns out to be $\kappa = 1.238$. In Fig.~\ref{fig:ColoronDecayWidth} we plot the dependence of the LO and NLO decay width on the cutoff scale $\Lambda$, that was introduced through the radiative mass splitting between the KK-quark and gluon. 

The calculation has been performed using publicly available computing tools sublimated by in-house routines. To generate the necessary diagrams and amplitudes we incorporated the MUED Feynman rules into {\sc FeynArts 3}~\cite{feynarts}. The color, Dirac and Lorentz algebra was performed with {\sc FeynCalc}~\cite{feyncalc}.

To simplify the treatment of tensor loop integrals, the one-loop amplitude was
contracted with the Born amplitude and the sum over the spins of external
particles carried out before any tensor reduction. As a result, most tensor
structures in the numerator of the loop integrand can be cancelled against
propagator denominators. For the remaining tensor integrals, Passarino-Veltman
reduction has been used \cite{pv}. One thus arrives at a final result in terms
of standard one-loop basis functions. A similar procedure has been employed for the calculations in the following sections. The IR-divergent basis integrals necessary for the Coloron decay were taken from Ref.~\cite{Hopker:1996sx}.


\section{Contribution of higher KK modes to KK-gluon decay}
\label{sc:ued}

 When calculating the contributions from modes with $n\geq 2$ to the decay $G_1 \to Q_1\bar{q}$, one needs to sum up diagrams with KK modes of level $n$ in the loops up to some order $N$, both in the vertices and in the counterterms. $N$ will later be identified with the cutoff scale $\Lambda R$.
The wave function renormalization is performed in analogy to the previous section, but now including level-$n$ modes in the loops. Then the remaining UV divergence is absorbed into a \msbar coupling renormalization of the form
\begin{align}
\delta Z_g = -\frac{\gs^2}{32\pi^2}\,
  \biggl (\frac{1}{\epsilon} - \gamma_{\rm E} + \log(4\pi) \biggr )
 \sum_{n=2}^N \beta_n\,,
\end{align}
from which one then can extract the contribution $\beta_n$ of the level to the overall beta function
\begin{align}
\frac{\partial \as^{-1}}{\partial\log{\mu}} = -\frac{1}{\as^2} \frac{\partial \as}{\partial\log{\mu}} =  \sum_n \frac{\beta_n}{2\pi}.
\end{align}

\subsection{Standard Model Vertices}
\label{SMVertexcoeff}

\begin{figure}[t]
\vspace{1em}
\centering
\captionsetup{width=\linewidth}
\includegraphics[width=5in]{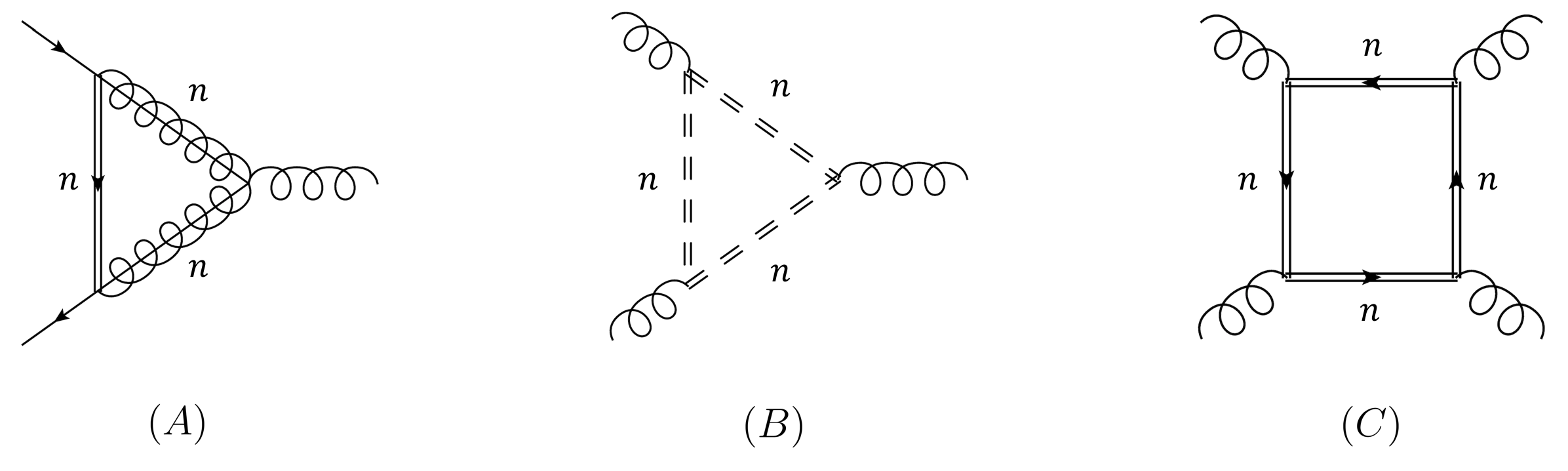}
\vspace{-1ex}
\mycaption{Sample diagrams for the KK-contributions to the SM vertices. (A) Three-point gluon Vertex (B) Quark-gluon Vertex (C) Four-point gluon vertex. Similar to Fig.~\ref{fig:ColoronDecaydiag}, a spring--straight line, solid double line and broken double line indicate the propagator of a level-n KK gluon, KK quark, and Goldstone mode of a KK gluon, respectively. 
\label{fig:SMVertex}}
\end{figure}

While we are ultimately interested in the decay of the KK-gluon, we will first study the impact of level-$n$ KK modes on the SM QCD vertices, in order to understand the renormalization procedure. There are three basic QCD vertices, the quark-gluon vertex, the three-gluon vertex and the four-gluon vertex. Example diagrams for each are shown in Fig.~\ref{fig:SMVertex}. All modes within in the loop have the same mode-number starting at $n=0$, as a consequence of KK-parity.

The standard result for the SM QCD beta function is reproduced by the diagrams with zero modes ($n=0$) at one loop and is given by
\begin{align}
\beta_\textrm{SM} = \left(\frac{11}{3}C_A - \frac{4}{3}n_qT_f\right).
\end{align}
When allowing KK modes with $n \geq 1$ in the loop, an additional contribution to the beta function is generated each time when one crosses a mass threshold $\frac{n}{R}$
\begin{align}
\beta_{n} = \left(\frac{7}{2}C_A - \frac{8}{3}n_qT_f\right).
\label{eq:betaN}
\end{align}
These contributions to the beta function lead to a running of the strong coupling constant that deviates from the Standard Model behavior, described for a renormalization scale between $\Lambda \leq \mu \leq \Lambda + R^{-1}$ by
\begin{align}
\alpha_S^{-1}\left(\mu\right) &= \alpha_S^{-1}\left(M_z\right) + \frac{\beta_\textrm{SM}}{2\pi} \log{\frac{\mu}{M_Z}} + \frac{\beta_N}{2\pi} \sum_{n=1}^{\Lambda R}\log{\frac{\mu}{nR^{-1}}} \nonumber\\
&=  \alpha_S^{-1}\left(M_z\right) + \frac{\beta_\textrm{SM}}{2\pi} \log{\frac{\mu}{M_Z}} + \frac{\beta_N}{2\pi}\left[\Lambda R \log{\frac{\mu}{R^{-1}}} - \log{\left(\Lambda R\right)!}\right],
\label{eq:levelnrunning}
\end{align}
where the sum over KK-modes extends up to the UV-cutoff scale $\Lambda$.
It has to be noted that the sum is strictly speaking only defined for an integer cutoff number, i.e.\ $\Lambda R$ in \eqref{eq:levelnrunning} is understood as the argument of the Gauss floor function $\lfloor{\Lambda R} \rfloor$, rounding down to the largest integer not exceeding $\Lambda R$.\\

Since we take all external legs to be massless there is no dipole operator being generated in the Lagrangian. We find that the vector coupling $gq\bar{q}$ as well as the three and four-point gluon couplings receive the same Wilson coefficient $C_\textrm{SM}$. Summing over all modes within the loop up to the cutoff scale that coefficient then reads
\begin{align}
C_\textrm{SM}\left(\mu\right) = \frac{\gs^3}{192\pi^2}\sum_{n=1}^{\Lambda R}\left[2C_A-(21C_A - 16n_qT_f)\log{\frac{n^2}{(\mu R)^2}}\right].
\label{eq:SMCoeff}
\end{align}
Notice the that the first term in the bracket is independent of the renormalization group running and represents a threshold correction which is not contained in the beta function.

For large cutoff scales, and setting $\mu=\Lambda$,  the Wilson coefficient has a leading linear dependence on $\Lambda$ and can be well approximated by
\begin{align}
C_\textrm{SM}\left(\Lambda\right) = \frac{\gs^3}{192\pi^2}\left[4\left(11C_A-8n_qT_f\right)\Lambda R - \left(21C_A - 16n_qT_f\right)\log{\Lambda R}\right] + \mathcal{O}\left(\frac{1}{\Lambda R}\right).
\label{eq:SMapprox}
\end{align}
For the computational details of the approximation see appendix~\ref{ap:AppendixA}.

\subsection{KK-Vertices}
\label{KKVertexcoeff}
\begin{figure}[t]
\vspace{1em}
\centering
\captionsetup{width=\linewidth}
\includegraphics[width=5in]{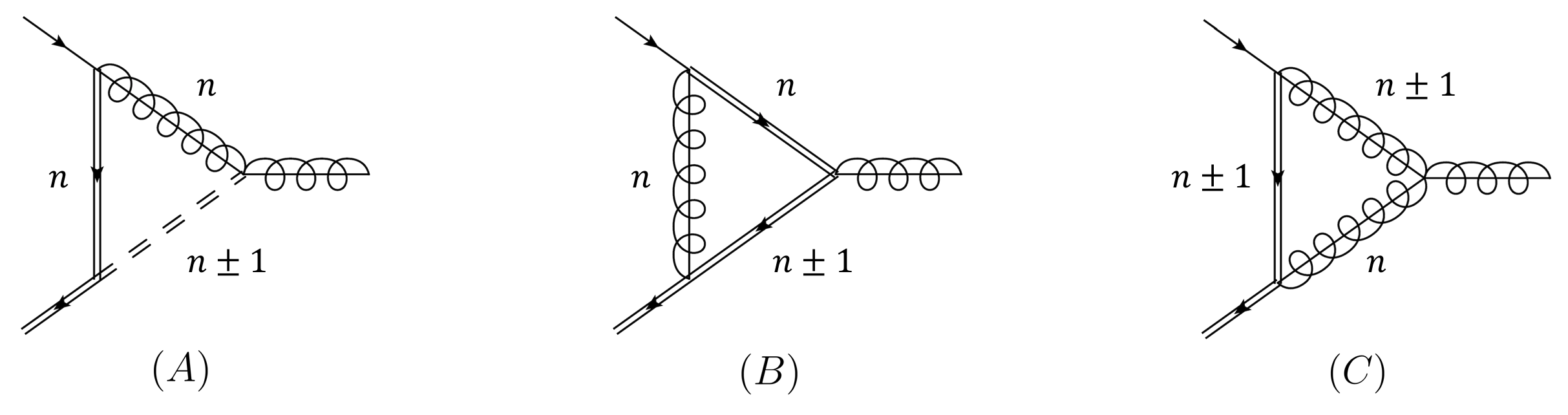}
\vspace{-1ex}
\mycaption{Sample diagrams for the KK-contributions to the vertex with two level-1 modes as external legs.
\label{fig:KKVertex}}
\end{figure}

In the case of two external legs being KK-particles the one-loop corrections to the vertex induce both a vertex-like and dipole operator and the total interaction under consideration becomes
\begin{align}
 -i C_{qQ_1G_1} \gamma^\mu P_L - D_{qQ_1G_1} \frac{\sigma^{\mu\nu}}{2R^{-1}}q_\nu P_L\,,
 \label{eq:interactionqQG}
\end{align}
where $q$ is the KK-gluon momentum.
If only SM modes and the first KK-mode are allowed in the loop, the coupling renormalization of eq.~\eqref{eq:Vertexrenorm} contributes to the beta function with $\beta_\textrm{Coloron}$.
The contributions to the vertex function from higher modes are shown in Fig.~\ref{fig:KKVertex} and can be straightforwardly calculated for every new mode allowed in the loop. The renormalization procedure is again done level by level and requires the on-shell field renormalizations for the external legs.

The beta function resulting from the coupling renormalization at every level $n\geq 2$ is identical to the one we found for the SM vertices for $n\geq 1$, reported in eq.~\eqref{eq:betaN}.

For the Wilson coefficient of the vector coupling of eq.~\eqref{eq:interactionqQG} we separate the NLO contribution into terms proportional to $C_A, C_F$ and $n_q$ and find 
\begin{align}
C_\textrm{KK} \left(\mu\right) =\; &-\frac{\gs^3}{64\pi^2}C_F\sum_{n=1}^{\Lambda R}\left[10 + (1+5n)\log{\frac{n^2}{(n+1)^2}}\right]\\
 &+ \frac{\gs^3}{64\pi^2}C_A\sum_{n=1}^{\Lambda R}\left[15+4n+4n^2+n\left(8+3n+2n^2\right)\log{\frac{n^2}{(n+1)^2}}-7\log{\frac{(n+1)^2}{(\mu R)^2}}\right]\nonumber\\
 &-\frac{\gs^3}{12\pi^2}n_qT_f\sum_{n=1}^{\Lambda R}\left[\frac{13}{6}+2n+2n^2+\frac{n}{2}\left(3+3n+2n^2\right)\log{\frac{n^2}{(n+1)^2}}-\log{\frac{(n+1)^2}{(\mu R)^2}}\right] \nonumber
\label{eq:KKCoeff}
\end{align}
Additionally we can calculate the dipole coefficient, which is finite and does not require renormalization:
\begin{align}
D_\textrm{KK} = \frac{\gs^2}{64\pi^2}\left(C_A^2 - 4\right)\left(C_A-2C_F\right)\sum_{n=1}^{\Lambda R}\left[2+\frac{1+2n}{2}\log{\frac{n^2}{(n+1)^2}}\right].
\end{align}
As before we set $\mu=\Lambda$ and can expand the Wilson coefficient in the limit of large $\Lambda$ and find
\begin{align}
C_\textrm{KK}\left(\Lambda\right) = \frac{\gs^3}{192\pi^2}\left[4\left(11C_A-8n_qT_f\right)\Lambda R - \left(42C_A - 32n_qT_f + 9 C_F\right)\log{\Lambda R}\right] + \mathcal{O}\left(\frac{1}{\Lambda R}\right).
\label{eq:KKapprox}
\end{align}
So both the coefficient of the SM operators and the KK operators show the same leading order behavior, although the $\log \Lambda R$ terms differ between eqs.~\eqref{eq:SMapprox} and \eqref{eq:KKapprox}.
Additionally we find for the dipole moment
\begin{align}
D_{qQ_1G_1} = \frac{\gs^2}{64\pi^2}(C_A^2-4)(C_A-2C_F)\left[\log({2\pi})-2\right] + \mathcal{O}\left(\frac{1}{\Lambda R}\right).
\end{align}
The dipole operator therefore is well-defined even in the limit $\Lambda \rightarrow \infty$.

\subsection{Flow Equation Analysis}
The previous results can be put into perspective through a renormalization flow analysis of the 5D uncompactified theory. Our solution strategy of the one-loop integro-differential equation is a non-local heat kernel expansion and closely follows Ref.~\cite{Codello:2015oqa}.

The effective action of a theory in saddle-point approximation can schematically be written as 
\begin{align}
\Gamma_k = S - \frac{1}{2}\textrm{Tr}\left[\log\left(\frac{\delta^2 S}{\delta\Phi \delta\Phi} + \mathcal{R}_k\right)\right],
\label{eq:saddlepoint}
\end{align}
where $S$ is the action and $\Phi$ represents a generic field.
We also introduced coarse graining through an explicit regulator function $\mathcal{R}_k$ designed to suppress the infrared modes of the theory. A standard choice is Litim's regulator \cite{Litim1}, which is given by
\begin{align}
\mathcal{R}_k(p^2) = \left(k^2 - p^2\right) \Theta\left(k^2 - p^2\right),
\label{eq:regulator}
\end{align}
in terms of the Heaviside theta function $\Theta$. It is optimized in the sense described in \cite{Litim2}.

Taking the derivative with respect to the renormalization time parameter $t = \log{\frac{k}{k_0}}$ yields the one loop flow equation, which can be written symbolically as
\begin{align}
\partial_t \Gamma_k = \frac{1}{2} \textrm{Tr}\left[\partial_t \mathcal{R}_k \left(\frac{\partial^2 S}{\partial\Phi \partial\Phi} + \mathcal{R}_k\right)^{-1}\right].
\label{eq:floweq}
\end{align}
This is the one-loop approximation can be obtained from the full functional renormalization group equation~\cite{Wettericheq}, by fixing the effective action to its UV limit $\Gamma_\Lambda = S$.\\
On the right-hand side of this flow equation we can identify a heat kernel $h_k$ which allows for a non-local expansion~\cite{NonLocalFE1,NonLocalFE2}. To lowest order and in the absence of gravity the expansion of the flow equation reduces to
\begin{align}
\partial_t \Gamma_k = \frac{1}{2\left(4\pi\right)^{\frac{D}{2}}}\int{d^Dx\left[\textrm{Tr}[\mathbb{1}]Q_{\frac{D}{2}}\left[h_k\right]+\textrm{Tr}\left[Ug_UU\right]+\textrm{Tr}\left[\Omega_{MN}g_\Omega\Omega^{MN}\right]\right]}.\label{eq:kernelexpansion}
\end{align}
where $\Omega_{MN}=\left[D_M,D_N\right]$ is the 5D gauge connection of the theory and $U = D^2 \mathbb{1} + \Delta$ the non-derivative part of the Laplacian, which carries all representations of the theory. Furthermore the expansion depends on coefficient functions given by
\begin{align}
g_U{(z,k)} &= \frac{1}{2}\int_0^1{d\xi\; Q_{\frac{D}{2}-1}\left[h_k^{z\xi(1-\xi)}\right]}\nonumber\\
g_\Omega{(z,k)} &=  \frac{1}{2z}Q_{\frac{D}{2}-1}\left[h_k\right] - \frac{1}{2z}\int_0^1{d\xi\; Q_{\frac{D}{2}-1}\left[h_k^{z\xi(1-\xi)}\right]},
\end{align}
which in turn can be expressed through the "Q-functionals" that take the form
\begin{align}
Q_n\left[h_k\right] &= \frac{1}{\Gamma\left[n\right]} \int_0^\infty{ds s^{n-1} h_k\left(s,w\right)} \;\;\;\;\; \textrm{for}\;\; n > 0 \nonumber\\
Q_{-n}\left[h_k\right] &= (-1)^n \frac{\partial^n}{\partial s^n} h_k\left(s,w\right)\bigg|_{s=0} \;\;\;\;\;\;\;\;\; \textrm{for} \;\; n \in \mathbb{Z} \leq 0,
\end{align} 
which can be found analytically for the optimized regulator \cite{NonLocalFE2}.

\paragraph{QED-like contributions:}
Let us begin by applying the flow equation eq.~\eqref{eq:floweq} to QED containing a massless fermion. From this one can extract the $T_f$ contribution for the QCD renormalization group flow in MUED. For this purpose, the fermion-gauge vertices are furnished with an extra factor $T^a$ each. After dismissing all operators that contain external fermions one obtains the heat kernel form of the flow equation
\begin{align}
\partial_t \Gamma_k = \frac{1}{2}\textrm{Tr}\left[\frac{\partial_t R_k(-\partial^2)^{MN}}{-\partial^2 \eta^{MN} + R_k(-\partial^2)^{MN}}\right]-\frac{1}{2}\textrm{Tr}\left[\frac{\partial_t R_k(\Delta)}{-D^2 + U + R_k(\Delta)}\right],
\end{align}
where the first term does not contain any fields and will therefore be ignored.
The non-derivative part $U$ of the Laplacian and the gauge connection $\Omega_{MN}$ are in this scenario given by
\begin{align}
U = -\frac{1}{2}\sigma^{MN}F_{MN}  \qquad \textrm{and} \qquad \Omega_{MN} =  iF_{MN}.
\end{align}
Armed with these tools we can apply the non-local heat kernel expansion of eq.~\eqref{eq:kernelexpansion} to the right hand side of the QED flow equation and find\footnote{To evaluate the Dirac trace note that $\left\{\Gamma_M,\Gamma_N\right\} = 2\eta_{MN}$,  $\textrm{Tr}\left[\Gamma^M \Gamma^N\right] = 4\eta^{MN}$ and $\textrm{Tr}\left[\Gamma^K \Gamma^L \Gamma^M \Gamma^N\right] = 4\left(\eta^{KL} \eta^{MN} - \eta^{KM} \eta^{LN} + \eta^{KN} \eta^{LM}\right)$ as well as $\Gamma_M \Gamma^M = 5$, whereas the Dynkin index $T_f$ arises when taking the color trace $\text{Tr}{\left[T^a,T^b\right]} = T_f \delta^{ab}$}
\begin{align}
\partial_t \Gamma_k = -\frac{T_f}{2}\frac{1}{(4\pi)^{\frac{D}{2}}}\int{d^Dx \left[\textrm{Tr}\left[\mathbb{1}\right]Q_{\frac{D}{2}}{\left[h_k\right]} + F_{MN}\left(2g_U - Dg_\Omega\right)F^{MN}\right]}.\label{eq:qedexpansion}
\end{align}
This form of the flow equation can now be compared to
\begin{align}
\Gamma_k\left[A\right] \bigg|_{F^2} = \int{d^Dx\; \frac{Z_k}{4}F_{MN}\left[1 + \Pi_k\left(z\right)\right]F^{MN}},
\end{align}
where the polarization function $\Pi_k(z)$ depends on $z=-D^2$.
This leads to an integro-differential equation for the wave function renormalization constant $Z_k$ and the polarization function
\begin{align}
\partial_t\left(Z_k(1+\Pi_k(z))\right) = -\frac{T_f}{(4\pi)^{D/2}}\int_0^1{d\xi \left\{ Q_{\frac{D}{2}-2}[h_k] + \frac{D}{2z}\left(Q_{\frac{D}{2}-1}\left[h_k\right] - Q_{\frac{D}{2}-1}\left[h^{z\xi(1-\xi)}_k\right]\right) \right\}}.\label{eq:fullqed}
\end{align}
To extract the beta function from this equation we consider the limit of zero momentum transfer, $z \rightarrow 0$, and exploit the fact that $\Pi_k(0)=0$. In that limit the second term of the equation behaves like
\begin{align}
\frac{1}{2z}\int_{0}^\infty{d\xi\; \left(Q_{\frac{D}{2}-1}\left[h_k\right]-Q_{\frac{D}{2}-1}\left[h_k^{z\xi(1-\xi)}\right]\right)} \xrightarrow{z\rightarrow 0} &\int_0^\infty{d\xi\; \frac{\xi(1-\xi)}{2}Q_{\frac{D}{2}-2}\left[h_k\right]}\nonumber\\
&= \frac{1}{12} Q_{\frac{D}{2}-2}\left[h_k\right].
\end{align}
We can now find the QED-like contributions to the anomalous dimension of the gluon field as well as the beta function, after identifying $\partial_t Z_k = 2\partial_t(g_k^{(5)})^{-1}$ for the 5D coupling, leading to
\begin{align}
\eta_k^\textrm{QED} = -\frac{\partial_t Z_k}{Z_k} = T_f \frac{g^2}{3\pi^2}kR  \qquad \textrm{and} \qquad \partial_k\as^{-1} =-\frac{2T_f}{3\pi}R,
\label{eq:Tfflow}
\end{align}
in terms of the 4D effective coupling $g = \sqrt{4\pi}\, g^{(5)}$. Note that the MUED model contains both a $SU(2)$ singlet and doublet quark for every flavor, so that an additional factor 2 must be included when comparing with section~\ref{SMVertexcoeff}.

\paragraph{Non-Abelian QCD contributions:} 
For QCD in $D$ dimensions one can proceed in a similar fashion. The contribution proportional to $C_A$ can be most conveniently extracted from the Yang-Mills part of QCD in MUED. The tree-level action in background field gauge reads 
\begin{align}
S_\textrm{QCD} = \int{d^Dx -\frac{1}{4}F^a_{MN}F^{a,MN} - \frac{1}{2\xi} \left(\overline{D}^{ab}_M \delta A^{a,M}\right)^2} - \Big(D^{ab}_M\overline{c}^b\Big)\left(\overline{D}^{M,ac}c^c\right),
\end{align}
where $\overline{D}_M$ denotes the gauge covariant derivative with respect to the background field, while $D_M$ contains the full field. $c_a$ is the corresponding Faddeev-Popov ghost.

The one-loop flow equation, applied to the above action, can be brought into the standard heat kernel form and is taken from Ref.~\cite{Codello:2013wxa}:
\begin{align}
\partial_t \Gamma{\left[\overline{A}\right]} = \frac{1}{2} \textrm{Tr}{\left[\frac{\partial_t \mathcal{R}_k{(\overline{D}_T)}-\eta^A_k\mathcal{R}_k{(\overline{D}_T)}}{\overline{D}_T + \mathcal{R}_k{(\overline{D}_T)}}\right]} - \textrm{Tr}{\left[\frac{\partial_t \mathcal{R}_k{(-\overline{D}^2)}-\eta^A_k\mathcal{R}_k{(-\overline{D}^2)}}{-\overline{D}^2 + \mathcal{R}_k{(-\overline{D}^2)}}\right]},
\end{align}
with the gauge covariant laplacian $D_{T,ab}^{MN} = -D_{ab}^2\eta^{MN} + U_{ab}^{MN}$.\\
From this definition we find the non-derivative part and the gauge connection in our conventions to be
\begin{align}
U_{MN}^{ab} = 2 f^{abc} F^c_{MN}  \qquad  \textrm{and} \qquad   \Omega_{MN}^{ab} = - f^{abc} F^c_{MN}.
\end{align}
After applying the non-local heat kernel expansion to the first term of the above flow equation and comparing it to the expected form of the effective action we are again left with
\begin{align}
\partial_t\left(Z_k(1+\Pi_k(z))\right) = \frac{C_A}{(4\pi)^{D/2}}\int_0^1 d\xi\; \biggl\{ &4Q_{\frac{D}{2}-2}\left[h_k^{z\xi(1-\xi)}\right] \notag \\ &+\frac{D-2}{z}\left(Q_{\frac{D}{2}-1}\left[h_k^{z\xi(1-\xi)}\right]-Q_{\frac{D}{2}-1}\left[h_k\right]\right)\biggr\}.
\end{align}
When taking the limit $z\rightarrow 0$ one finds the non-Abelian QCD contribution to the anomalous dimension of the gluon field as well as to the beta function
\begin{align}
\eta_k^\textrm{QCD} = -\frac{\partial_t Z_k}{Z_k} = -7C_A\frac{g^2}{16\pi^2} kR  \qquad \textrm{and} \qquad \partial_k\as^{-1} = \frac{7C_A}{4\pi}R.
\label{eq:CAflow}
\end{align}


\section{Phenomenological analysis}
\label{sc:pheno}

From eqs.~\eqref{eq:Tfflow} and \eqref{eq:CAflow} one finds that the cutoff dependence of the effective gauge coupling is given by
\begin{align}
\frac{1}{\gs(\Lambda)} - \frac{1}{\gs(0)}
 = \frac{\gs^3}{8\pi^2}\biggl( \frac{7}{4}C_A - \frac{4}{3}n_qT_f \biggr ) \Lambda R\,.
\label{eq:flow}
\end{align}
Since this was obtained in the framework of an uncompactified 5D theory, there is no distinction between couplings of zero modes and higher KK modes.

The result in eq.~\eqref{eq:flow} is numerically close, but not identical to the linear $\Lambda R$ terms in eqs.~\eqref{eq:SMapprox} and \eqref{eq:KKapprox}. The difference stems from threshold corrections that contribute to \eqref{eq:SMapprox} and \eqref{eq:KKapprox} at each KK level. Thus, even when we study the dependence of the effective vertex coupling on $\Lambda$ for $\Lambda R \gg 1$, i.e.\ at scales much beyond the compactification radius, there is a non-vanishing impact of the compactification. In other words, when considering MUED at large scales one does not trivially recover the uncompactified 5D theory.

\begin{figure}[t!]
\centering
\captionsetup{width=\linewidth}
\vspace{-1em}
\includegraphics[width=12cm]{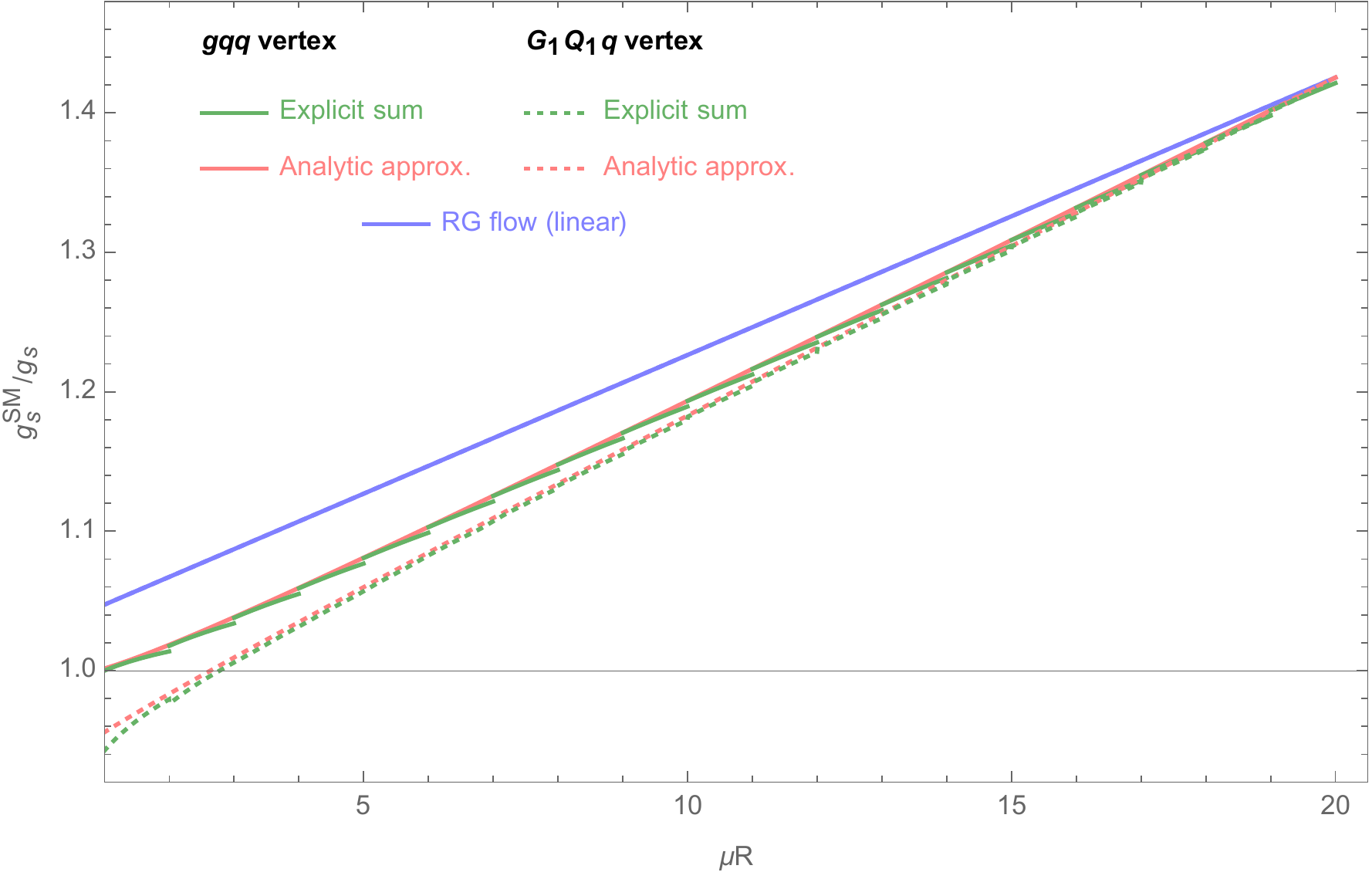}
\vspace{-1ex}
\mycaption{Running of the SM vertex function $q-q-g$ (solid line) and the KK vertex $q-Q_1-G_1$ (dashed line). We compare the full summation up to the cutoff with the asymptotic expansion as well as the results obtained from the FRGE running. The integration constants have been fixed such that the couplings coincide at  $\Lambda R=20$.}
\label{fig:Wilsoncompare}
\end{figure}
This is illustrated in Fig.~\ref{fig:Wilsoncompare}. In this plot it is assumed that the there is some fundamental 5D coupling at the scale $\Lambda_0 =20 R^{-1}$. The curves illustrate the effect of integrating out all energy scales beyond $\Lambda R^{-1}$ in different approximations. The solid blue line corresponds to the result \eqref{eq:flow} from the flow equation. The solid green curve depicts the result from the explicit calculation of the $gq\bar{q}$ vertex diagrams, as in eq.~\eqref{eq:SMCoeff} with $\mu=\Lambda$, whereas the solid orange curve is the approximation \eqref{eq:SMapprox}. The dotted green and orange curves are the equivalent cases for the $G_1Q_1\bar{q}$ vertex, eqs.~\eqref{eq:KKCoeff} and \eqref{eq:KKapprox}. The vertical axis is normalized such that the effective $gq\bar{q}$ coupling at $n=1$ is equal to its SM value.

First of all, the plot demonstrates that the large-$\Lambda R$ expansions (orange) provide an excellent approximation to the level-by-level calculations (green). Furthermore, it shows that all approaches lead to similar results for the dependence of the effective gauge couplings on the truncation scale. However, there are difference at the few-percent level, which stem from the following two facts: The flow equation is insensitive to threshold corrections from the compactification; and the $G_1Q_1\bar{q}$ and $gq\bar{q}$ vertices have different logarithmic dependencies on $\Lambda$. Note that the plot also includes the running from the coloron contribution, see eq.~\eqref{eq:Vertexrenorm}.

From a low-energy perspective, on the other hand, the dependence of the gauge vertices on the cutoff $\Lambda$ is not directly observable. However, if KK gluons and KK quarks should be discovered in the future, one can compare the strength of the SM $gq\bar{q}$ coupling with the $G_1Q_1\bar{q}$ vertex. While the leading linear $\Lambda R$ dependence drops out in this difference, there is still some sensitivity to the cutoff scale from the logarithmic terms in eqs.~\eqref{eq:SMapprox} and \eqref{eq:KKapprox}: 
\begin{align}
C_{qQ_1G_1}(\Lambda)-C_{qqg}(\Lambda) &=
 -\frac{\gs^3}{192\pi^2}(21C_A+9C_F-16n_qT_f) \log \Lambda R + \mathcal{O}\left(\frac{1}{\Lambda R}\right).
\end{align}
Including the contribution to the running from the lowest modes, see eq.~\eqref{eq:Vertexrenorm}, and assuming that the $gq\bar{q}$ and $G_1Q_1\bar{q}$ couplings are identical at the scale $\mu=\Lambda$, one finds that the observable couplings depends on $\Lambda$ as follows:
\begin{align}
\frac{g_{qQ_1G_1}(R^{-1})}{g_{qqg}(R^{-1})} &\approx \frac{23\gs^2}{192\pi^2} \log \Lambda R\,.
\end{align}
For $\Lambda R = 10 \dots 50$, this amounts to an effect between 3.5\% and 6\%.

This means that the prediction of the decay rate for $G_1 \to Q_1 \bar{q}$, including $\mathcal{O}(\as)$ corrections, depends on the unknown UV completion of MUED. At the same time, the impact of this UV sensitivity is rather mild, at the level of a few percent, which may be negligible for most practical purposes.

While a more complete analysis of different processes in MUED and larger classes
of extra-dimensional models is beyond the scope of this work, we expect that a
similar conclusion can be reached in these cases. This conjecture is based on
our observation that the leading UV sensitivity cancels when normalizing the KK
gauge boson vertex to the SM gauge boson vertex, as dictated by the
renormalization flow equation. The cancellation still works when including the
threshold corrections at each KK level, and thus it should hold in any
extra-dimensional extension of the SM. The next-to-leading term, while enhanced
by  $\log(\Lambda R)$, is nevertheless numerically rather modest.

\black


\section{Conclusions}
\label{sc:concl}

Models with extra dimensions generally feature infinite KK towers of new states that need to be considered when calculating loop corrections within those models. KK-number conserving operators receive corrections from the full spectrum of these modes. The corrections are in general dependent on a cutoff scale $\Lambda$ at which the model breaks down and some unspecified UV completion is required to describe the physics. In this article we investigated the numerical impact of this unknown cutoff parameter on physical observables. As a concrete framework we considered QCD in a spacetime with one additional universal extra dimension, which is compactified on a circle with a $\mathbb{Z}_2$ orbifold. 

In the first half of this article, the problem was discussed from the viewpoint of the compactified 4D effective theory. We computed the full one-loop QCD corrections to the SM gauge-boson vertex $gq\bar{q}$ as well as the vertex $G_1Q_1\bar{q}$ (or, equivalently, $G_1\bar{Q}_1q$), which contains two level-1 KK modes. These include vertex diagrams, on-shell counterterms for the external legs, and \msbar coupling counterterms generated by loops up to KK level $n$. The \msbar scale dependence of the $gq\bar{q}$ vertex can be described through the regular QCD beta function, where the well-known SM result is supplemented by an extra term each time one of the KK thresholds is crossed. Additionally, the Wilson coefficient of the $gq/\bar{q}$ vertex receives a finite threshold correction from each KK level, which is not described by the beta function.

For the $G_1Q_1\bar{q}$ ($G_1\bar{Q}_1q$) vertex we proceeded similarly. For concreteness, we considered the physical process $G_1\rightarrow \bar{q} Q_1$. The NLO corrections to this decay exhibit soft divergencies in the virtual vertex contributions, which can be cancelled against the real radiation contributions with the two cutoff phase-space slicing method. This part of the calculation is identical to a 4D coloron model. Overall, the beta function for this vertex differs from the SM $gq/\bar{q}$ vertex, but the contribution from higher KK modes ($n>1$) is identical for the $gq\bar{q}$ and $G_1Q_1\bar{q}$ beta functions. Furthermore, the Wilson coefficient for the $G_1Q_1\bar{q}$ vertex receives KK threshold corrections, which differ from the ones found for the SM vertex.

The results for both vertices can be conveniently written in terms of an expansion for large values of the cutoff scale $\Lambda$. It turns out that their leading terms, which are linear in $\Lambda R$, are identical. The first difference between the two vertices appears at the subleading order $\log{(\Lambda R)}$. Higher orders beyond the log term are numerically very small and can be neglected.

In the second half, we studied the cutoff dependence within uncompactified 5D QCD, using the functional renormalization group flow equation. To solve the equation and extract the beta function we applied the saddle-point approximation and utilized the non-local heat kernel expansion method. Through this approach we were able to find the flow equation analogue of the one-loop beta function. Its coefficient coincides with the leading order behavior of the running of the coupling found for the explicit diagrammatic calculation in the compactified theory. However, the uncompactified 5D framework is not able to reproduce the contributions from the threshold corrections in the 4D framework.

When comparing the diagrammatic 4D calculations of the $gq\bar{q}$ and $G_1Q_1\bar{q}$ vertices, as well as the 5D flow equation result, one finds that the NLO prediction for the decay $G_1\rightarrow \bar{q} Q_1$ is indeed sensitive to the choice of the UV cutoff and thus to the unknown high-scale physics. However, the numerical impact of this uncertainty is numerically rather modest, since the leading contributions cancel in the comparison.


\section*{Acknowledgments}

\noindent
This work has been supported in part by the National Science Foundation under
grant no.\ PHY-1519175.


\appendix

\section{Asymptotic Expansion}
\label{ap:AppendixA}
The analytic summation using elliptical Jacobi theta functions, that has been employed in Ref.~\cite{original,Dienes:1998vg,Kubo:1999ua}, is not an accurate method for our case, since we are defining our coupling renormalization within the \msbar scheme. We are however able to recover the same results in the large $\Lambda R$ limit, as the theta function method does within the on-shell scheme, using a somewhat different approach.\\
We begin by performing the summation up to mode $N = \left\lfloor {\Lambda R}\right\rfloor$ analytically. The only non-trivial sums appearing are of the form
\begin{align}
\sum_{n=1}^{N}{\log{\frac{n}{n+1}}} =\log{\frac{\Gamma[N+1]}{\Gamma[N+2]}}
\end{align}
which can be rewritten using the Euler $\Gamma$ function as a generalization of the factorial, and 
\begin{align}
\sum_{n=1}^{N}{n\log{\frac{n}{n+1}}} = \log{\Gamma[N+2]} + \zeta'[-1, N+1] - \zeta'[-1,N+2],
\end{align}
which can be carried out using the Riemann $\zeta$ function. The prime denotes a derivative with respect to the first argument of the generalized $\zeta$ function $\zeta[s,a] = \sum_{n=0}^{\infty}(n+a)^{-s}$. One solves the sums involving higher moments analogously, like e.q.:
\begin{align}
&\sum_{n=1}^{N}{n^2 \log{\frac{n}{n+1}}} =\nonumber\\
&= -\frac{1}{6} + 2\log{A} -\log{\Gamma[N+2]} + \zeta'[-2,N+1] - \zeta'[-2,N+2] + 2\zeta'[-1,N+2] ,
\end{align}
where $A = 1.282427\cdots$ denotes the Glaisher-Kinkelin constant.\\
\\
For the asymptotic expansion of the first sum, in a region where $\Lambda R$ becomes large we simply utilize Stirling's Formula
\begin{align}
\log{\Gamma[N]} = N(\log{N} - 1) -\frac{1}{2}\log{\frac{N}{2\pi}} + \frac{1}{12N} +\mathcal{O}\left(N^{-2}\right).
\end{align}
For the sums involving higher powers of $n$, the asymptotic expansion of the derivative of the generalized Riemann $\zeta$ function can be performed according to
\begin{align}
\zeta'[-1,N] &= -\frac{N^2}{4}(1-2\log{N}) -\frac{N}{2}\log{N} +\frac{1}{12}(1+\log{N}) +\frac{1}{720N^2} + \mathcal{O}\left(N^{-3}\right)\nonumber\\
\zeta'[-2,N] &= -\frac{N^3}{9}(1-3\log{N}) -\frac{N^2}{2}\log{N} +\frac{N}{12}(1+2\log{N}) -\frac{1}{360N}+ \mathcal{O}\left(N^{-3}\right)
\end{align}
It is worth noting at this point that the Riemann $\zeta$ function is related to the elliptical Jacobi theta function through
\begin{align}
\int_0^\infty{\frac{dt}{t}\,(\vartheta_n(it) -1) t^{s/2}} = \frac{2}{\pi^{\frac{s}{2}}} \Gamma\left[\frac{s}{2}\right] \zeta\left[s,1\right],
\end{align}
making our order by order recovery of the results derived with the aide of the theta function not entirely surprising.


\end{document}